\documentclass[traditabstract]{aa}
\usepackage{graphicx}
\usepackage{txfonts}
\begin{document}
\headnote{Research Note}
\title{Stability of the toroidal magnetic field in rotating
       stars} 
\author{
A.~Bonanno\inst{1,2}, V.~Urpin\inst{1,3,4}
}
%\offprints{}
\institute{
           $^{1)}$ INAF, Osservatorio Astrofisico di Catania,
           Via S.Sofia 78, 95123 Catania, Italy \\
           $^{2)}$ INFN, Sezione di Catania, Via S.Sofia 72,
           95123 Catania, Italy \\
           $^{3)}$ A.F.Ioffe Institute of Physics and Technology,
           194021 St.Petersburg, Russia \\ 
           $^{4)}$ Isaac Newton Institute of Chile, Branch in St. Petersburg,
           194021 St. Petersburg, Russia
}
\date{\today}

\abstract{The magnetic field in stellar radiation zones can play an important role in phenomena such as mixing, angular momentum transport, etc.
We study the effect of rotation on the stability of a predominantly toroidal magnetic field in the radiation zone. In particular we  
considered the stability in spherical geometry by means of a linear analysis in the Boussinesq approximation. 
It is found that the effect of rotation on the stability depends on a magnetic configuration.
If the toroidal field increases with the spherical radius, the instability cannot be suppressed entirely even by a very fast rotation. Rotation can only
decrease the growth rate of instability. If the field decreases with the radius, the instability has a threshold and can be completey suppressed.}

\keywords{instabilities - magnetohydrodynamics - stars: interiors - stars: magnetic field - Sun: interior}

%\keywords{stars: interiors - stars: magnetic field - Sun: interior}

\maketitle

\section{Introduction}  
It is rather uncertain which magnetic field can be present in stellar  
radiation zones but this field can play an important role in many 
phenomena in stars such as mixing, angular momentum transport, 
formation of tachocline etc. (see, e.g., Gough \& McIntyre 1998, Heger 
et al. 2005, Eggenberger et al. 2005, Mathis \& Zahn 2005). Likely, 
dynamos cannot operate in radiation zones, where no strong flows are 
available to sustain a vigorous dynamo action. Possibly, relic magnetic 
fields acquired by the star at the early stage of evolution can persist 
there. This type of fields could have formed, for instance, because of 
differential rotation that could have stretched the lines of a weak 
primordial seed field into a dominant toroidal field. Since the conductivity 
is high, the large-scale relic field could survive in the radiation 
zone during the life-time of a star. 

The magnetic field, however, can evolve in a radiation zone also 
because of the development of various instabilities. For example, 
the magnetorotational instability could occur if the radiation zone
is magnetized and rotates differentially. However, differential
rotation is unlikely in radiation zones and, perhaps, can exist only
in stellar tachoclines. Over the past decade, instabilities of the
stellar tachocline have been extensively studied (see, e.g., Dikpati
et al.(2009) and reference therein). The tachocline is thin and its
stability properties are rather peculiar. For instance, Gilman \&
Fox (1997) showed that the tachocline latitudinal shear is unstable
to nonaxisymmetric disturbances when a toroidal magnetic field is
present. Instabilities in the tachocline have been studied in detail
by Dikpati et al.(2009) for a wide range of rotation and toroidal 
field profiles. Since rotation is rigid in radiation zones, 
instabilities of the magnetic field most likely are current-driven. 
These instabilities do not require differential rotation, and they 
are well studied in cylindrical geometry in the context of laboratory 
fusion research (see, e.g., Freidberg 1973, Goedboed 1971, Goedbloed 
\& Hagebeuk 1972, Tayler 1973a,b, 1980). In astrophysical conditions, 
the instability caused by electric currents might have a number of 
characteristic features even in cylindrical geometry (see Bonanno 
\& Urpin 2008a,b, 2011). The nonlinear evolution of the Tayler 
instability was studied by Bonanno et al. (2012), who argued that 
symmetry-breaking can give rise to a saturated state with nonzero 
helicity even if the initial state has zero helicity. A production 
of nonzero helicity is crucial for the dynamo action.    

The stability of the spherical magnetic configurations is studied in much 
less detail. This problem is of particular interest in relation
to Ap star magnetism (Braithwaite \& Spruit 2004). With numerical 
simulations Braithwaite \& Nordlund (2006) studied the stability of 
a random initial field. They argued that this field relaxes on a stable 
mixed magnetic configuration with both poloidal and toroidal components. 
A study of the magnetic configurations with a predominantly toroidal 
field is of particular importance for radiation zones because this 
field can be easily formed by differential rotation at the early 
evolutionary stage. Numerical modeling by Braithwaite (2006) confirmed 
that the toroidal field with $B_{\varphi} \propto s$ or $\propto s^2$ 
($s$ is the cylindrical radius) is unstable to the $m=1$ mode, as 
was predicted by Tayler (1973a). The stability of azimuthal fields 
has also been studied by Spruit (1999). The author used a heuristic 
approach to estimate the growth rate and criteria of instability. 
Unfortunately, many of these estimates and criteria are valid only
near the rotation axis and do not apply in the main fraction of the 
volume of a radiation zone where the stability properties can be 
qualitatively different (see Zahn et al. 2007). Recently,  Bonanno \& 
Urpin (2012) have considered the stability of the toroidal field in 
radiation zones by making use of a linear analysis and taking into 
account stratification and thermal conductivity. It is widely believed
that stratification can suppress the Tayler instability. Bonanno \&
Urpin (2012) calculated the growth rate of instability and argued 
that the stabilizing influence of gravity can never entirely suppress 
the instability caused by electric currents. However, a stable 
stratification can essentially decrease the growth rate of instability. 

In this paper, we consider the effect of rotation on the stability
of magnetic configurations with a predominantly toroidal field. 
Rotation is often considered as one more factor that can suppress the 
Tayler instability and stabilize the magnetic configurations. For 
instance, Spruit (1999) found that the growth rate of the Tayler 
instability in a rotating star should be on the order of $\sim 
\omega_A (\omega_A /\Omega)$ if $\Omega \gg \omega_A$, where $\omega_A$ 
and $\Omega$ are the Alfven frequency and angular velocity of the star, 
respectively. Stability of the toroidal field in rotating stars has 
been considered by Kitchatinov (2008), and Kitchatinov \& R\"udiger 
(2008), who argued that the magnetic instability is determined by the 
threshold field strength at which the instability sets. Estimating 
this threshold in the solar radiation zone, the authors imposed the 
upper limit on the magnetic field to be $\approx 600$ G. The stability of 
the toroidal field in a rotating radiation zone has been studied by 
Zahn et al. (2007) in the particular case $B_{\varphi} \propto s$. 
The particular type of oscillatory modes found by these authors is 
related to rotation and is stable in the nondissipative limit. 
However, instability can occur in the form of an oscillatory diffusive 
instability if dissipation is provided by radiative or Ohmic diffusion.

The paper is organized as follows. The basic equations and mathematical
formulation of the problem are presented in Sec.2. This is followed by 
results of numerical calculations of the growth rate and frequency of
the instability in Sec.3. The paper closes with a summary of the main
results and some remarks in Sec.4.

\section{Basic equations}
Consider the stability of an axisymmetric toroidal magnetic field in 
the radiation zone using a high conductivity limit. We work in spherical 
coordinates ($r$, $\theta$, $\varphi$) with the unit vectors 
($\vec{e}_{r}$, $\vec{e}_{\theta}$, $\vec{e}_{\varphi}$). We assume that 
the radiation zone rotates with the angular velocity $\vec{\Omega}$=const
and that the toroidal field depends on $r$ and $\theta$, $B_{\varphi}= 
B_{\varphi}(r, \theta)$. If the magnetic field is subthermal (so that the
magnetic pressure is lower than the gas pressure), one can 
apply the incompressible limit for a consideration of low-frequency modes. 
{We conduct the analysis of hydromagnetic modes in the rest frame 
rather than in the corotating frame. Generally, some rotation-related effects 
can be missing in this case. Some insights into how hydromagnetic 
waves behave in the corotating frame can be found in the papers by Hide 
(1969) and Acheson \& Hide (1973). For instance, effects due to the density 
stratification can be neglected when considering the dispersion relation 
for hydromagnetic oscillations if the Brunt-V\"ais\"al\"a frequency is 
much lower than twice the angular velocity. However, the Brunt-V\"ais\"al\"a 
frequency is basically much higher than the angular velocity in stellar 
radiation zones and these effects can be neglected. }

The MHD equations in the incompressible limit  read
\begin{eqnarray}
\frac{\partial \vec{v}}{\partial t} + (\vec{v} \cdot \nabla) \vec{v} = 
- \frac{\nabla p}{\rho} + \vec{g} 
+ \frac{1}{4 \pi \rho} (\nabla \times \vec{B}) \times \vec{B}, 
\end{eqnarray}
\begin{equation}
\frac{\partial \vec{B}}{\partial t} - \nabla \times (\vec{v} \times \vec{B}) 
= 0,
\end{equation}
\begin{equation}
\nabla \cdot \vec{v} = 0, \;\;\; \nabla \cdot \vec{B} = 0, 
\end{equation}
where $\vec{g}$ is gravity. 
The equation of thermal balance reads in the Boussinesq approximation
\begin{equation}
\frac{\partial T}{\partial t} + \vec{v} \cdot (\nabla T - \nabla_{ad}T) =
0,
\end{equation} 
where $\nabla_{ad} T$ is the adiabatic temperature gradient.

In the basic (unperturbed) state, the gas is assumed to be in hydrostatic 
equilibrium, then
\begin{equation}
\frac{\nabla p}{\rho} = \vec{g} + \frac{1}{4 \pi \rho} 
(\nabla \times \vec{B}) \times \vec{B} + \vec{e}_s \;\Omega^2 \;r \sin \theta,
\end{equation}
where $\vec{e}_s$ is the unit vector in the cylindrical radial direction.
The rotational energy is assumed to be much lower than the 
gravitational one, $g \gg r \Omega^2$. The origin and structure of 
the magnetic field in radiation zones are unknown. However, $\vec{g}$ is 
approximately radial in our basic state since we assume that  the magnetic 
energy is subthermal. Only small variations of the density and 
temperature are required in the meridional direction to balance the 
centrifugal and Lorentz forces for a given magnetic configuration. 

We consider a linear stability. Weak perturbations will be indicated by 
subscript 1, while unperturbed quantities will have no subscript. Linearizing 
Eqs.(1)-(4), we take into account that weak perturbations of the 
density and temperature in the Boussinesq approximation are related by
$\rho_1/\rho = -\beta (T_1/T)$, where $\beta$ is the thermal expansion 
coefficient. For weak perturbations, we use a local approximation in the 
$\theta$-direction and assume that their dependence on $\theta$ is 
proportional to $\exp( - i l \theta)$, where $l \gg 1$ is the longitudinal 
wavenumber. Since the basic state is stationary and axisymmetric, the 
dependence of perturbations on $t$ and $\varphi$ can be taken in the 
exponential form as well. Then, perturbations are proportional to 
$\exp{(\sigma t - i l \theta - i m \varphi)}$, where $m$ is the azimuthal 
wavenumber. The corresponding wavevectors are $k_{\theta}=l/r$ and 
$k_{\varphi}=m/r \sin \theta$, respectively. The dependence on $r$ should be 
determined from Eqs.(1)-(4). 

The problem of the magnetic field stability is complicated from 
both the physical and computational points of view. The conclusions of 
different studies are often contradictory (compare, for example, the 
results by Spruit (1999) and Kitchatinov \& Ruediger (2008)), and
very often the authors do not discuss in detail the reason of the controversy.
Because of the complexity of this problem, in our opinion, the best approach 
is to separately consider the role played by different physical factors 
(gravity, rotation, conductivity, etc.) in modifying the global structure of the unstable modes.
Only after this process has been clarified is it  possible to 
arrive at a global picture of the parameter space.
The main motivation of this paper  is therefore to clarify the effect of rotation
by considering different radial profiles of the toroidal field, in contrast to the 
study proposed  in detail by Bonanno \& Urpin (2012), where only the effect of 
gravity was considered.
Therefore, we consider a simplified problem assuming that 
stratification is neutral and $\nabla T = \nabla_{ad}T$. In this case,
the stabilizing effect of gravity is neglected and we can study how the
instability is affected by rotation alone. 
For the sake of simplicity, we also 
assume that the unperturbed density is approximately homogeneous in the 
radiation zone. As we mentioned above, meridional variations of the
density should be small in a subthermal magnetic field. A radial variation
is not small in real stars but  omitting it does not qualitatively change the main 
conclusions and substantially simplifies calculations. 
Eliminating all variables in favor of $v_{1r}$, we obtain with the accuracy 
in terms of the lowest order in $(k_{\theta} r)^{-1}$ 
\begin{eqnarray}
(\sigma_0^2 + \omega_A^2 + D \Omega_i^2) \; v_{1r}'' +
\left( \frac{4}{r} \sigma_0^2 + \frac{2}{H} \omega_A^2  \right) v_{1r}'
\\ \nonumber 
+ \left[ \frac{2}{r^2} \sigma_0^2
- k_{\perp}^2 (\sigma_0^2 +  \omega_A^2 ) - D \Omega_e^2 k_{\theta}^2 +
\right.
\\ \nonumber
\left.
\frac{2}{r} \omega_A^2 \left( \frac{1}{H} 
\frac{k_{\perp}^2}{k_{\varphi}^2} - \frac{2}{r} 
\frac{k_{\theta}^2}{k_{\varphi}^2} D \right) - 
i \sigma_0 \Omega_e \left( \frac{k_{\varphi}}{r} +
4 D \frac{k_{\theta}^2}{r k_{\varphi}} \frac{\omega_A^2}{\sigma_0^2}
\right) \right] v_{1r} = 0,
\end{eqnarray}
where the prime denotes a derivative with respect to $r$ and
\begin{eqnarray}
\sigma_0 = \sigma - i m \Omega, \;\;\; 
\omega_A^2 = \frac{k_{\varphi}^2 B_{\varphi}^2}{4 \pi \rho}, \;\;\;
D = \frac{\sigma_0^2}{\sigma_0^2 + \omega_A^2}, 
\\ \nonumber
\Omega_i=2 \Omega \cos \theta,\;\;\; \Omega_e=2 \Omega \sin \theta,\;\;\; 
k_{\perp}^2 = k_{\theta}^2 + k_{\varphi}^2, \;\;\;
\\ \nonumber
\frac{1}{H} = \frac{\partial}{\partial r} \ln (r B_{\varphi}). 
\end{eqnarray}
If $\Omega = 0$, Eq.(6) transforms into the equation derived
by Bonanno \& Urpin (2012).

Some important stability properties of the toroidal field can be derived 
directly from Eq.(6). Consider perturbations with a very short radial 
wavelength for which one can use a local approximation in the radial 
direction, such as $v_{1r} \propto \exp( -ik_{r} r)$, where $k_{r}$ is 
the radial wavevector. If $k_{r} \gg \max(k_{\theta}, k_{\varphi})$, 
then Eq.(6) yields with the accuracy in terms of the lowest order in 
$(k_{r} r)^{-1}$ the following dispersion equation
\begin{equation}
\sigma_0^2 + \omega_A^2 + D \Omega_i^2 = 0, 
\end{equation}
or
\begin{equation}
\sigma_0^4 + \sigma_0^2 (2 \omega_A^2 + \Omega_i^2) + \omega_A^4 
= 0.
\end{equation}
It is easy to show that this equation has only imaginary roots. Therefore, 
modes with a short radial wavelength are always stable to the current-driven 
instability in contrast to the result obtained by Kichatinov (2008) and 
Kichatinov \& R\"{u}diger (2008).

\section{Numerical results}
We assume that the radiation zone is located at $R_i \leq r \leq R$ or,
introducing the dimensionless radius $x= r/R$, at $x_i \leq x \leq 1$
where $x_i= R_i/R$. We choose the internal radius of the radiation zone, 
$x_i$, to be equal to $0.1$ from computational reasons. We have verified 
that our results are basically insensitive to the precise value of $x_i$ 
as long as it is close to the center. 

The toroidal field can be represented as
\begin{equation}
B_{\varphi} = B_0 \psi(x) \sin \theta,
\end{equation}
where $B_0$ is the characteristic field strength and $\psi \sim 1$ is a 
function of the spherical radius alone. The dependence of $\psi$ on $x$ 
is uncertain in the radiation zone and, in this paper, we consider 
three different possibilities. Fig.~1 plots the profiles $\psi(x)$ for the
models (1), (2), and (3) used in our calculations. The case where 
the field reaches its maximum at the outer boundary (model (2)) can 
mimic, for example, the radiation zone of a star with a convective 
envelope. In this case, the bottom of a convection zone likely is  the 
location of the toroidal field generated by a dynamo action. The 
toroidal field can penetrate into the radiation zone, for instance, 
because of diffusion. Model (2) can also mimic the toroidal field 
in the liquid core of neutron stars. Likely, the magnetic field of these 
objects is generated by turbulent dynamo during the very early phase of 
evolution when the neutron star is subject to hydrodynamical 
instabilities (see Bonanno et al. 2005, 2006). A large-scale dynamo is 
most efficient in the surface layers where the density gradient is 
highest. Therefore, the generated field increases outward and reaches 
its maximum in the outer layers (Bonanno et al. 2005, 2006). This 
magnetic field can be subject to current-driven instabilities 
after the end of the initial phase. Note that dynamo induced by turbulent 
motions generates not only a large-scale field but also small-scale fields 
of complex topology (Urpin \& Gil 2004). The stability properties of the 
configurations with mixed small- and large-scale fields can be very 
particular and are not studied yet. The model (3) with a decreasing 
toroidal field can mimic a star whose magnetic field is generated in the 
inner convective core. Generally, all three models (1), (2), and (3) can be 
representative of the stars with relic magnetic fields because 
the details of the formation of these fields are very uncertain.

\begin{figure}
\includegraphics[width=9cm]{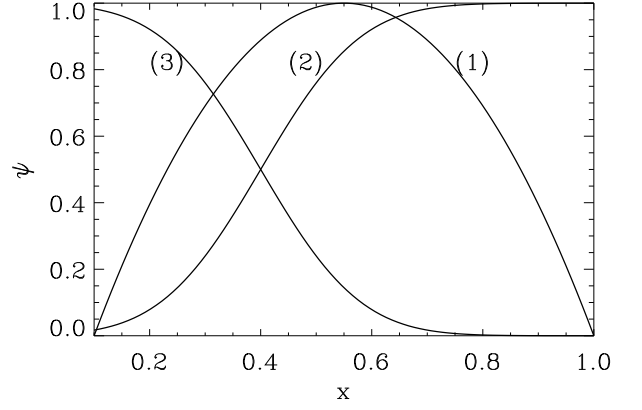}
\caption{Dependence of the toroidal field on the spherical radius
for models (1), (2), and (3).}
%\label{angle}
\end{figure}
 
Introducing the dimensionless quantities as
\begin{equation}
 \Gamma = \frac{\sigma_0}{\omega_{A0}}, \;\;\; 
\eta = \frac{2 \Omega}{\omega_{A0}}, 
\end{equation} 
where $\omega_{A0}^2 = B_{0}^2 /4 \pi \rho R^2$, we can transform Eq.(6) 
into a dimensionless form. This equation with the corresponding boundary 
conditions describes the stability problem as a nonlinear eigenvalue 
problem. Fortunately, the main qualitative features of this problem are 
not sensitive to the choice of boundary conditions. That is why we choose 
the simplest boundary conditions and assume that $v_{1r}= 0$ at $r=R_i$ 
and $r=R$. Generally, solutions of Eq.(6) are complex. It is more 
convenient to split all quantities into the real and imaginary parts and 
to solve numerically the set of two real coupled equations that follows 
from Eq.(6). Note that coefficients of Eq.(6) and the corresponding 
dimensionless equation depend on $\theta$, which in turn leads to the dependence 
of $\Gamma$ on it. 

The stability properties in the spherical geometry are qualitetively
different from those in the cylindrical geometry (see Bonanno \& 
Urpin 2012). Therefore, the results obtained, for instance, for the 
toroidal field dependending on the cylindrical radius alone (see Spruit 
1999, Zhang et al. 2003) does not apply to more general magnetic 
configurations. The stability problem becomes particularly complex 
if the radiation zone rotates.

In Fig.~2, we plot the growth rate and frequency of the Tayler modes
as functions of $\eta$ at different $\theta$ for the model (2). Like
the case of a non-rotating star, the Tayler instability is the most
efficient at the equator (see also Bonanno \& Urpin 2012). At small
$\eta$, the growth rate is of the order of $1.4 \omega_{A0}$ but it 
clearly shows some suppression for a faster rotation. Suppression 
becomes significant already at relatively low values of $\eta \sim
2-3$. The growth rate decreases with an increase of $\eta$ approximately 
as $1/ \eta$ and it does not vanish even at very large $\eta$. A
similar behavior was obtained by Spruit (1999), who considered
the instability near the rotation axis for $B_{\varphi} = 
B_{\varphi}(s)$. It turns out that rotation can never entirely suppress the 
Tayler instability of  model (2)  but only decreases the 
growth rate. Tayler modes are oscillatory in a rotating 
radiation zone in contrast to the nonrotating case. The frequency is 
basically comparable to the growth rate and also decreases when the rotation 
becomes faster.     
  
\begin{figure}
\includegraphics[width=9cm]{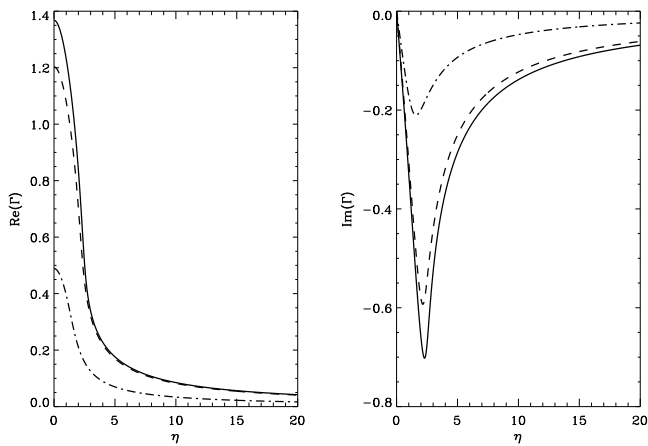}
\caption{Growth rate (left panel) and frequency (right panel) 
of the Tayler modes as functions of the rotational parameter $\eta$
for $\theta = \pi/2$ (solid), $\pi/6$ (dashed), and $\pi/10$
(dash-and-dotted) and for  model (1). The longitudinal and 
azimuthal wavenumbers are $l = 10$ and $m=1$, respectively.}
%\label{angle}
\end{figure}

Fig.~3 shows the results of calculations of the growth rate and frequency
for model (1). The results are very similar to those obtained for 
 model (2). As in the previous case, the instability is most efficient 
at the equator and its growth rate decreases when it approaches the rotation axis. 
This result is at variance with the widely accepted opinion that toroidal 
magnetic configurations are always unstable at the axis (see, e.g., Spruit 
1999). This opinion is usually based on the similarity of the spherical 
magnetic configuration near the axis and the axisymmetric cylindrical 
configuration. However, this analogy is generally incorrect because in 
spherical geometry, the toroidal field near the axis also depends on 
the radial coordinate along the axis. Therefore, it can be unjustified to 
apply the results obtained for a cylinder with $B_{\varphi}=B_{\varphi}(s)$ 
to the case of plasma near the symmetry axis in spherical geometry (Bonanno 
\& Urpin 2012). Indeed, in the latter case stability can crucially depend 
on the profile of the toroidal field along the symmetry axis. Like 
model (2), the instability of model (1) is strongly suppressed 
by rotation. Suppression becomes essential at $\eta \geq 3-4$. At large 
$\eta$, the growth rate decreases $\propto 1/\eta$, as in the previous case. 
The stabilization cannot be reached at any large $\eta$ but the growth rate 
can be substantially reduced.  
\begin{figure}
\includegraphics[width=9cm]{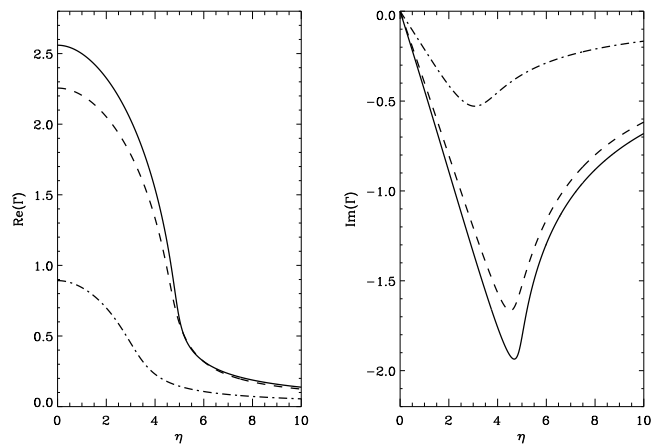}
\caption{Same as in Fig.2 but for model (2).}
%\label{angle}
\end{figure}

In Fig.~4, we plot the growth rate and frequency as functions of $\eta$
for  model 3. Magnetic configurations with a rapidly decreasing toroidal 
field are stable in a cylindrical geometry. For example, stability 
properties of the toroidal field are determined by the parameter $\alpha = 
d \ln B_{\varphi}(s) / d \ln s$. The field is unstable to 
axisymmetric perturbations if $\alpha  > 1$ and to nonaxisymmetric 
perturbations if $\alpha> - 1/2$ (Tayler 1973a,b, 1980). The situation 
is qualitatively different in  spherical geometry and even the 
toroidal field, decreasing rapidly with $r$, can be unstable. Again,
the instability is most efficient at the equator. However, its efficiency
decreases drastically toward the pole and the instability does not occur
in the region around the rotation axis. The critical angle, $\theta_{cr}$,
which distinguishes the stable and unstable regions, is $\sim 40^{\circ}$.
At variance with the models (1) and (2), the instability of the model (3) 
is determined by the threshold field strength. The threshold is not very
high and corresponds to $\eta \approx 7$. Therefore, the Tayler instability 
is entirely suppressed in model (3) if $\Omega \gtrsim 3.5 \omega_{A0}$.
The growth rate is vanishing everywhere in the radiation zone for more
rapidly rotating stars.   
 \begin{figure}
\includegraphics[width=9cm]{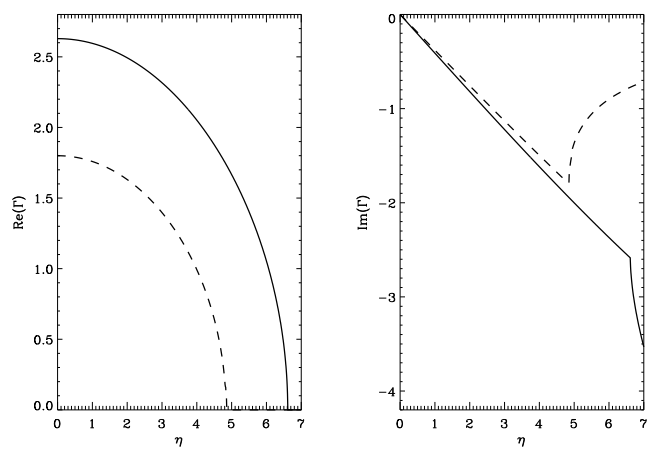}
\caption{Same as in Fig.2 but for model (3) and $\theta = \pi/2$ 
(solid) and $\pi/3$ (dashed).}
%\label{angle}
\end{figure}

\section{Conclusion}
We have considered the stability of the toroidal field in rotating 
stellar radiation zones. The stability properties of the spherical magnetic 
configurations turn out to be qualitatively different from those of the 
cylindrical configurations (Bonanno \& Urpin 2012). For instance, in 
contrast to widely accepted opinion, the toroidal field can be 
stable even near the symmetry axis. The reason for this behavior is the 
spatial dependence of the magnetic field along the axis which can provide 
a stabilizing effect. Therefore, a direct analogy between the stability of a 
cylinder with the azimuthal field and the toroidal field near the axis 
in radiation zones is generally incorrect.
 
Rotation provides a stabilizing effect on the Tayler instability. The
effect of rotation can be characterized by the parameter $\eta$, which
is generally large in radiation zones. It turns out that the effect of 
rotation depends critically on the magnetic configuration. If the toroidal 
field increases with the spherical radius within the radiation zone (or 
some fraction of it, see  models (1) and (2)) then rotation cannot entirely
suppress the Tayler instability  even at very large $\eta$. The 
instability growth rate is discernible for any rotation but  can 
be substantially reduced at very large $\eta$. A reduction of the growth 
rate becomes important even at a relatively low angular velocity $\Omega
\sim 2-3 \omega_{A0}$. At large $\eta$, the growth rate behaves like 
$\omega_{A0} (\omega_{A0} / \Omega)$, as was obtained by Spruit (1999) 
for a particular case $B_{\varphi} = B_{\varphi}(s)$ near the rotation 
axis. The Alfven timescale, $\omega_{A0}^{-1}$, is 
short compared to the stellar life-time, therefore even a suppressed 
instability with a reduced growth rate can be significant for radiation 
zones. It should be noted also that, most likely, the field does not decay 
to zero because of this instability. When the field becomes weaker, the 
growth rate of the instability decreases and the field cannot decay to 
values lower than those resulting from the condition that the growth rate 
is on the order of the inverse life-time of a star. Therefore,  a weak 
field can change only insignificantly during the life of the star, 
although its radial profile can be unstable.

If the toroidal field decreases with the spherical radius (see Fig.~4 for
model (3)), the rotation effect is different. In contrast to 
models (1) and (2), the instability of  model (3) is determined by the 
threshold field strength. The threshold is not very high and corresponds 
to $\eta \approx 7$. Therefore, the Tayler instability is entirely suppressed 
in the model (3) if $\Omega \gtrsim 3.5 \omega_{A0}$ and modes are stable 
everywhere in the radiation zone for more rapidly rotating stars. Higher 
eigenmodes are suppressed more strongly than the fundamental one and perturbations 
with a short radial wavelength are always stable. Since instability is not
suppressed at $\eta < 7$, this implies that the magnetic field should satisfy
the condition $B_0 \gtrsim 0.3 \Omega R \sqrt{ 4 \pi \rho}$. Estimating 
$\Omega R \sim 2 \times 10^5$ cm/s and $\rho \sim 0.1$ g/cm$^3$, we obtain 
that instability can arise in the radiation zone of the Sun if $B_0 \gtrsim
7 \times 10^4$ G. This estimate is more than two orders of magnitude 
higher than
that obtained by Kichatinov \& R\"udiger (2008). These authors considered 
stability of the toroidal field assuming that perturbations are global
in the meridional direction and short-scaled in radius. As a result, they
obtained that the most rapidly growing modes modes are indefinitely short
in the radial direction if diffusion is neglected. This conclusion is questionable
because simple analytic considerations (see Section 2 of the present
paper) show that perturbations with short radial wavelength should always be
stable. In particular our numerical calculations also show that the fundamental eigenmode
has a higher growth rate than higher eigenmodes.   

Finally, we calculated the growth rate assuming a neutral
stratification of the radiation zone. If the stratification is stable, gravity
provides an additional stabilizing influence on the Tayler instability, as
was argued by Bonanno \& Urpin (2012). Therefore, suppression of the
instability can be even more pronounced and the calculated growth rates 
are likely to be only upper limits in real, magnetized stellar interiors.

\vspace{0.5cm}
\noindent
{\it Acknowledgments}.
VU acknowledges support from the European Science Foundation (ESF) within
the framework of the ESF activity "The New Physics of Compact Stars".
VU also thanks  the Russian Academy of Sciences for financial support under the
program OFN-15 and the INAF-Ossevatorio Astrofisico di Catania for hospitality.

{}

\end{document}